\documentclass{article}
\usepackage{hyperref}
\usepackage{graphicx}
\usepackage{natbib}
\usepackage[margin=0.5in]{geometry}

\begin{document}

\title{A5-miseq: an updated pipeline to assemble microbial genomes from Illumina MiSeq data}
\author{David Coil\thanks{Genome Center, University of California Davis, USA}, Guillaume Jospin,$^*$ and Aaron E. Darling\thanks{ithree institute, University of Technology Syndey, Australia, to whom correspondence should be addressed: aaron.darling@uts.edu.au}}

\maketitle

\begin{abstract}

\section{Motivation:}
Open-source bacterial genome assembly remains inaccessible to many biologists due to its complexity. 
Few software solutions exist that are capable of automating all steps in the process of \textit{de novo} genome assembly from Illumina data.

\section{Results:}
A5-miseq can produce high quality microbial genome assemblies from as little as 20-fold sequence data coverage on a laptop computer without any parameter tuning.
A5-miseq does this by automating the process of adapter trimming, quality filtering, error correction, contig and scaffold generation, and detection of misassemblies.
Unlike the original A5 pipeline, A5-miseq can use long reads from the Illumina MiSeq, use read pairing information during contig generation, and includes several improvements to read trimming.
Together these changes result in substantially improved assemblies that recover a more complete set of reference genes than previous methods.

\section{Availability:}
A5-miseq is licensed under the GPL open source license. Source code and precompiled binaries for Mac OS X 10.6+ and Linux 2.6.15+ are available from \url{http://sourceforge.net/projects/ngopt}

\section{Contact:} \href{aaron.darling@uts.edu.au}{aaron.darling@uts.edu.au}
\end{abstract}

\section{Introduction}

Genome assembly involves an entire data processing workflow starting with raw sequence data and ending with scaffolded contigs.  
The steps often consist of adapter trimming, quality filtering, error correction, creation of contigs, verification of contigs by mapping reads to the assembly, and the creation/verification of scaffolds.  

We previously published A5, a pipeline that automated all the steps to generate bacterial genome assemblies from raw Illumina data~\citep{Tritt12}.  
The workflow included five steps and the parameters for each were optimized on assemblies of Halophilic archaea and tested on \textit{E. coli}.

Since the publication of A5, Illumina's chemistry has advanced significantly and the MiSeq instruments are now capable of producing reads in excess of 400nt long, which is fourfold longer than what was previously possible on a HiSeq 2000.
The original A5 could not process reads longer than 150nt.
The longer reads make it possible to assemble genomes from less data overall, but doing so required major revisions to the data processing algorithms in A5. 

\begin{figure}[!tpb]
\centerline{\includegraphics[width=5.5in]{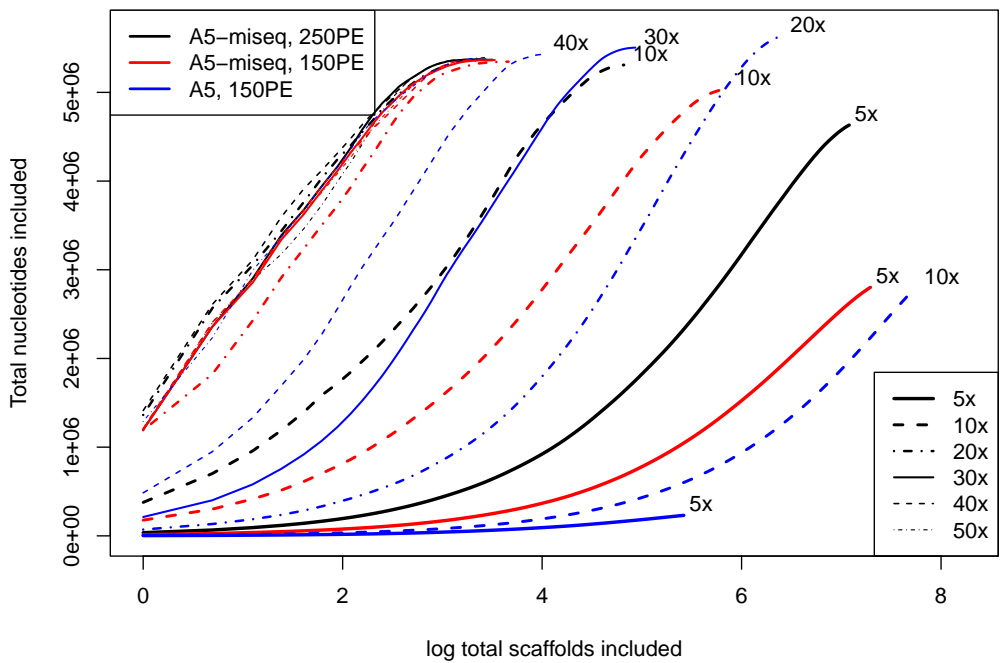}}
\caption{Contiguity of assemblies of the \textit{B. cereus} data from GAGE-B with increasing depths of coverage. A5 was run on reads cut down from 250nt to 150nt. A5-miseq was run on both the cut down reads (red) and the full length 250nt reads (black). A5-miseq assemblies become highly contiguous with 20x coverage, whereas A5 (blue) requires 50x coverage to achieve comparable contiguity.}\label{fig:dilution}
\end{figure}

We introduce a revised pipeline called A5-miseq which replaces several components of the original A5 pipeline with new software modules and produces substantially improved assemblies.

\begin{table*}[!t]
\scriptsize
\caption{Comparison of assembly accuracy between A5-miseq and other assemblers}
\begin{tabular}{ll|ccccc|ccccc|cccc}
\hline
& & \multicolumn{5}{ c| }{A5-miseq} & \multicolumn{5}{ c }{A5} & \multicolumn{4}{ c }{GAGE-B} \\
Organism                & Size & Frac & NGA50 & MA & MM & Genes & Frac & NGA50 & MA & MM & Genes & Frac & NGA50 & MA & Genes \\\hline
\textit{B. cereus}      & 5.43 & 99.9 & 486.8 & 7 & 1.4 & 5734 & 99.8 & 488.3 & 12 & 14.3 & 5669 & 99.9 (S) & 456 (sdn) & 1 (sdn) & 5439 (M) \\
\textit{R. sphaeroides} & 4.60 & 99.9 & 130.5 & 6 & 2.5 & 4426 & 99.6 & 146.9 & 19 & 10.3 & 4325 & 99.9 (S) & 151.8 (S) & 0 (sga) & 3562 (S) \\
\textit{M. abscessus}   & 5.09 & 99.4 & 232.5 & 12 & 3.1 & 4922 & 99.3 & 109.7 & 9 & 1.8 & 4873 & 99.4 (S) & 215.4 (S) & 3 (A) & 4361 (S) \\
\textit{V. cholerae}    & 4.03 & 99.6 & 196.7 & 15 & 5.1 & 3645 & 98.9 & 67.8 & 22 & 4.4 & 3510 & 99.6 (S) & 246.6 (S) & 3 (A) & 3564 (S) \\\hline
\end{tabular}
\label{Tab:01}
{Assembly accuracy for the A5-miseq and A5 pipelines measured on raw 100x coverage MiSeq PE250 GAGE-B data. Accuracy was measured by QUAST. Frac is the fraction of the reference genome represented in assembly scaffolds. NGA50: the N50 after breaking contigs at misassemblies. MA: number of misassemblies, MM: number of nucleotide mismatches per 100kbp, Genes: number of intact full-length genes recovered. For GAGE-B results, the single best assembly result for each metric and genome reported in \citet{Magoc13} is shown, with the assembler producing the best result indicated as follows: S=SPAdes 2.3, sdn=SOAPdenovo, M=MaSuRCA, A=ABySS, sga=SGA.}
\end{table*}

\section{Methods}

The A5-miseq pipeline consists of five steps: \textbf{(1) Read cleaning} -- sequence adapters and low quality regions are removed by Trimmomatic~\citep{Lohse12}. 
Errors in the reads are then corrected using SGA's $k$-mer based error correction algorithm~\citep{Simpson11}.
\textbf{(2) Contig assembly} -- paired and unpaired reads are used for assembly with the IDBA-UD algorithm~\citep{Peng12}. 
\textbf{(3) Crude scaffolding} -- contigs are scaffolded with any available large insert libraries using permissive parameters. \textbf{(4) Misassembly correction} -- misassemblies are detected on the basis of read pairs that do not map within the expected distance. Contigs and scaffolds found to contain misassemblies are broken. \textbf{(5) Final scaffolding} -- a final round of scaffolding with stringent parameters repairs any previously broken contigs. Assembly summary statistics and basecall quality estimates are also produced in stage 5.

A5-miseq substantially revises steps (1) and (2) relative to A5. 
In step (1) A5 would discard entire reads found to contain any amount of adapter readthrough. 
The standard Nextera XT protocol results in libraries where a large fraction of the reads ($>$50\% in extreme cases) contain adapter readthrough when sequenced with the currently standard paired-end 300nt read chemistry.
Instead of discarding such reads, only the contaminated portion of the read gets trimmed.

In step (2) A5-miseq employs a contig assembly algorithm (implemented in the IDBA-UD software) that exploits read pairing information during contig generation.
This improvement reduces the frequency with which misassembled contigs are formed during the contig generation step.
Thus fewer misassemblies must be detected and corrected in step (4) of the pipeline.
IDBA-UD required extensive source code revision to operate on Mac OS X and reduce its memory usage for laptop hardware.
These changes are available in the A5-miseq source code repository.

We benchmarked A5-miseq 20140521 and A5 on the raw GAGE-B MiSeq datasets.
GAGE-B includes paired-end 250nt MiSeq reads at 100x coverage for four organisms. 
We ran A5-miseq and A5 assemblies and obtained results for other assemblers from the GAGE-B publication~\citep{Magoc13}.
The original A5 can not assemble 250nt reads, so we used Trimmomatic to cut the reads down to 150nt (discarding 100nt) prior to assembly.
Running time and peak memory were measured with \texttt{/usr/bin/time -v} on a 2012 MacBook Air running Ubuntu 13.10.

To evaluate assembly accuracy we ran QUAST v2.2~\citep{Gurevich13} with the following parameters: \texttt{quast.py --gage -u -G ref.gff -R ref.fa ass.fa}, where \texttt{ref.fa} is the reference assembly from NCBI, \texttt{ref.gff} are the reference's annotated genes from NCBI, and \texttt{ass.fa} is the assembly. 
The \texttt{-u} option causes the genome fraction to be calculated in the manner used by QUAST v1.3. 
This yields results that are comparable to those in the GAGE-B paper.

In a separate experiment, we evaluated how the completeness of assemblies produced by A5 and A5-miseq changes in response to decreasing amounts of sequence data. 
To do so we took the 100x coverage GAGE-B datasets and randomly downsampled them to 50x, 40x, 30x, 20x, 10x, and 5x coverage, assembled them, and measured assembly accuracy with QUAST.

\section{Discussion}

The benchmarking results for A5-miseq on the GAGE-B data (shown in Table~\ref{Tab:01}) indicate that it offers substantial improvements over the original A5 pipeline.
A5-miseq assemblies contain more full length reference genes than A5 or any of the GAGE-B assemblers.
Relative to the original A5, the number of genes missing from A5-miseq assemblies is reduced by 1.6 to 3.8-fold on the GAGE-B datasets.
%
In most cases, A5-miseq assemblies have higher NGA50 values, fewer misassemblies, and fewer base calling errors than the original A5 pipeline.
A5-miseq produces results that are competitive with the best achieved by other assemblers on the GAGE-B data, but requires only minimal user-effort.

A5-miseq can recover nearly complete genome assemblies with much less sequence data than its predecessor. 
Figure~\ref{fig:dilution} shows that to recover a fixed fraction of the genome, e.g. 95\%, A5-miseq requires about half as much sequence data.
Relative to A5, A5-miseq achieves higher degrees of contiguity with less data.
This permits deeper multiplexing of sequencing experiments.

A5-miseq is computationally efficient. Assembly of the \textit{B. cereus} GAGE-B data completed in 2.2 hours with a peak memory usage of 4GB and 5.7GB disk usage on a laptop. Compute requirements for other bacterial genomes are similar.

One limitation of the GAGE-B data is that following its publication, assembly pipelines might be inadvertently tuned to produce high scores specifically on that dataset.
This could result in artificially high scores that do not accurately reflect the expected performance on other datasets. 
We assert that we have not tuned A5-miseq in any way to improve scores on the GAGE-B dataset, and to our knowledge, none of the component programs used by A5-miseq have been optimized for GAGE-B.

\section{Conclusion}

Genome assembly is a fast evolving field and software has been advancing rapidly. 
Although A5-miseq produces assemblies that are competitive with results in a recently published assembler evaluation~\citep{Magoc13}, it is likely that versions of other software that are currently under peer-review (e.g. SPAdes 3.0) might produce even better results.
SPAdes 2.3 automates many of the same steps that A5-miseq automates~\citep{Bankevich12} and in general produces excellent assemblies (Table~\ref{Tab:01}), with A5-miseq's main advantages being automated adapter trimming, more full length genes assembled, NCBI-ready outputs, and production of base call quality scores.
If possible, researchers interested in genome assembly should become acquainted with the various algorithms available before selecting a particular approach.  
A5-miseq should be particurlarly useful for researchers with limited bioinformatics experience or computing resources.

\section*{Acknowledgement}
We would like to thank users of the A5 pipeline for reporting bugs and usability problems with the software, in particular Piklu Roy Chowdhury and the manuscript peer reviewers.

\paragraph{Funding:} This work was supported in part by a collaborative agreement with the NSW Department of Primary Industries.

%
%

\end{document}